\documentclass[floats,aps,showpacs,twocolumn,superscriptaddress,groupedaddress]{revtex4}

\usepackage[dvips]{graphicx}
\usepackage{amsfonts}
\usepackage{amsmath,amssymb}
\usepackage{dsfont}
\usepackage{placeins}
\usepackage{afterpage}
\usepackage{flafter}
\usepackage{color}

\newcommand{\leftexp}[2]{{\vphantom{#2}}_{#1}{#2}}

\newcommand{\ud}{\text{d}}
\newcommand{\ui}{\text{i}}
\newcommand{\ue}{\text{e}}

\newcommand{\psiregm}{|m\rangle_{0}}

\newcommand{\psiregmr}{|m+r\rangle_{0}}
\newcommand{\psiregmpl}{\leftexp{0}{\langle} m'|}
\newcommand{\psiregmpr}{|m'+r\rangle_{0}}
\newcommand{\psiregmtr}{|m+2r\rangle_{0}}
\newcommand{\psiregmm}{|m-1\rangle_{0}}
\newcommand{\psich}{\psi_{\text{ch}}}

\newcommand{\Ureg}{U_{\text{reg}}}
\newcommand{\Hres}{H_{\text{res}}}

\newcommand{\Vrs}{\mathcal{V}_{\text{r:s}}}
\newcommand{\Ors}{\Omega_{\text{r:s}}}

\newcommand{\tVrs}{\bar{\mathcal{V}}_{\text{r:s}}}
\newcommand{\Irs}{I_{\text{r:s}}}
\newcommand{\mrs}{m_{\text{r:s}}}
\newcommand{\hpeak}{h_{\text{peak}}}
\newcommand{\hres}{h_{\text{res}}}
\newcommand{\gammad}{\gamma^{\text{d}}}
\newcommand{\A}{\mathcal{A}}
\newcommand{\Ars}{\mathcal{A}_{\text{r:s}}}
\newcommand{\dArs}{\Delta \mathcal{A}_{\text{r:s}}}

\begin{document}

\title{Regular-to-Chaotic Tunneling Rates: From the Quantum to the Semiclassical Regime}

\author{Steffen L\"ock}
\author{Arnd B\"acker}
\author{Roland Ketzmerick}
\affiliation{Institut f\"ur Theoretische Physik, Technische Universit\"at
             Dresden, 01062 Dresden, Germany}
\author{Peter Schlagheck}
\affiliation{Institut f\"ur Theoretische Physik, Universit\"at Regensburg, 93053 Regensburg, Germany}
\affiliation{Division of Mathematical Physics, Lund Institute of Technology, PBox 118, 22100 Lund, Sweden}

\date{\today}

\begin{abstract}
We derive a prediction of dynamical tunneling rates from regular to chaotic
phase-space regions combining the direct regular-to-chaotic tunneling mechanism
in the quantum regime with an improved resonance-assisted tunneling
theory in the semiclassical regime.
We give a qualitative recipe for identifying the relevance of nonlinear resonances in
a given $\hbar$-regime. For systems with one or multiple
dominant resonances we find excellent agreement to numerics.
\end{abstract}
\pacs{05.45.Mt, 03.65.Sq, 03.65.Xp}

\maketitle
\noindent

In mixed regular-chaotic systems the quantitative understanding of dynamical 
tunneling, which refers to classically forbidden transitions between
phase-space regions that are separated by dynamical barriers
\cite{DavHel1981}, represents one of the most challenging open problems in
semiclassical physics.
In the early 1990s it was found \cite{LinBal1990,BohTomUll1993,TomUll1994} 
that tunneling rates between 
phase-space regions of regular motion are substantially enhanced by the 
presence of chaotic motion, which was termed chaos-assisted tunneling.
Such dynamical tunneling processes are ubiquitous 
in molecular physics \cite{DavHel1981} and were realized with cold 
atoms in periodically modulated optical lattices
\cite{SteOskRai2001,HenHafBroHecHelMcKMilPhiRolRubUpc2001}.

The fundamental process in this context is tunneling between states in a 
regular island and the surrounding chaotic region.
In the quantum regime, $h \lesssim\A$, where Planck's constant $h$ is 
smaller but comparable to the area of the regular island $\A$, this 
process is dominated by a \textit{direct regular-to-chaotic
tunneling} mechanism \cite{HanOttAnt1984,PodNar2003,SheFisGuaReb2006}. 
An approach to determine
these direct tunneling rates was recently given in Ref.~\cite{BaeKetLoeSch2008}. 
It relies on a fictitious integrable system that resembles the regular 
dynamics within the island under consideration, and has been applied to
quantum maps \cite{BaeKetLoeSch2008}, billiard systems 
\cite{BaeKetLoeRobVidHoeKuhSto2008}, and the annular microcavity 
\cite{BaeKetLoeWieHen2009}. In the semiclassical regime, $h\ll \A$, however,
the direct tunneling contribution alone is incapable to describe the observed 
tunneling rates.

\begin{figure}[b]
  \begin{center}
    \includegraphics[angle = 0, width = 85mm]{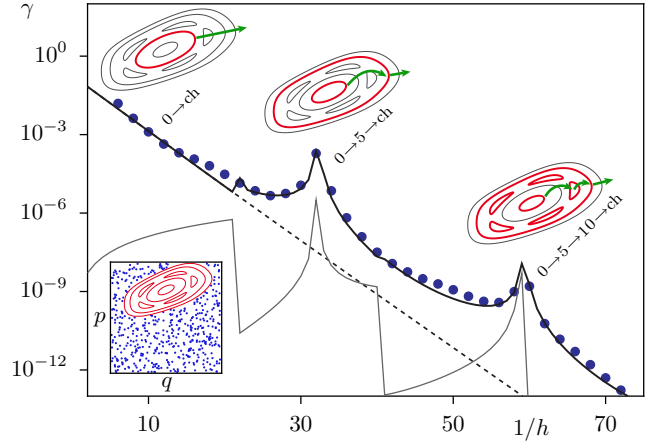}
    \caption{(Color online) Dynamical tunneling rates $\gamma$ from the
             innermost quantized torus ($m=0$) of a regular island
             with one dominant $5$:$1$ resonance. Numerical data (dots)
             is compared to the prediction of Eq.~\eqref{eq:rat_tunneling_rate}
             (solid line) and previous results in the quantum regime
             due to the direct tunneling process
             (dashed line, Eq.~\eqref{eq:gamma0}, Ref.~\cite{BaeKetLoeSch2008})
             and in the semiclassical regime due to resonance-assisted
             tunneling (gray solid line, Ref.~\cite{EltSch2005}).
             The lower inset shows the phase space of the 
             system \cite{RATMAPADD}. The upper insets illustrate the dominant 
             tunneling steps for three values of $h$.
            }
    \label{fig:results_one_resonance}
  \end{center}
\end{figure}

The existence of nonlinear resonances inside the regular island leads to 
peaks and plateaus in the tunneling rates. These can be 
qualitatively predicted by the theory of \textit{resonance-assisted tunneling} 
\cite{BroSchUllcomb,EltSch2005} as shown for quantum
maps \cite{EltSch2005}, periodically driven systems 
\cite{MouEltSch2006,WimSchEltBuc2006}, quantum accelerator modes 
\cite{SheFisGuaReb2006}, and for multidimensional molecular 
systems \cite{Kes2005,Kes2005b}. Quantitatively, however, deviations of 
several orders of magnitude appear
(see Fig.~\ref{fig:results_one_resonance}, gray line). 
Hence, both mechanisms alone give no reliable quantitative prediction of 
tunneling rates in generic systems.

In this Letter we derive a unified framework that combines the direct 
regular-to-chaotic and the resonance-assisted tunneling mechanisms.
Beyond the direct tunneling contribution there are additional contributions:  
They consist of resonance-assisted tunneling steps within the regular island 
and, in contrast to previous studies \cite{EltSch2005,SheFisGuaReb2006},
a final direct tunneling step to the chaotic sea; see insets in 
Fig.~\ref{fig:results_one_resonance}. This leads to an excellent prediction
of tunneling rates from the quantum to the semiclassical regime.
In particular, it includes the $\hbar$-regime, which is relevant 
for the search of experimental signatures of resonance-assisted tunneling.

Specifically, we consider one-dimensional kicked systems, described by the 
classical Hamiltonian $H(p,q,t)=T(p)+ V(q)\sum_{k\in\mathbb{Z}} \delta(t-k)$ 
and by the quantum time-evolution operator 
$U=\exp[-\ui V(q)/\hbar]\exp[-\ui T(p)/\hbar]$, which exhibit one major
regular island embedded in the chaotic sea. 
This mixed phase-space structure gives rise to eigenstates of $U$ that are
mainly regular or chaotic, i.e.\ concentrated on a torus inside the regular 
region or spread out over the chaotic sea. However, they do have small 
components in the other region of phase space. Consequently, a wave packet 
started on the $m$th quantized torus
($m=0,1,\dots,m_{\text{max}}-1$) of the island will, in the presence of  
absorbing boundary conditions in the chaotic sea, 
leak out from the island $\sim \ue^{-\gamma_m t}$ with a 
characteristic tunneling rate $\gamma_m$. 
It describes the mean coupling to the chaotic sea, where fluctuations 
from individual chaotic states are averaged.

We now consider the presence of a prominent nonlinear $r$:$s$ resonance 
($s$ oscillations match $r$ driving periods) within the island. 
For small $\hbar$, this resonance gives rise to couplings between 
different regular states \cite{BroSchUllcomb}, which compete
with direct tunneling and lead to additional pathways into the chaotic 
sea. We derive (see below) the tunneling rate of the 
$m$th quantized state as
\begin{equation}
\label{eq:rat_tunneling_rate}
  \gamma_m = \gammad_{m}+ |A_{m,r}^{(r:s)}|^2 \gammad_{m+r} +
             |A_{m,r}^{(r:s)} A_{m,2r}^{(r:s)}|^2 \gammad_{m+2r} + \dots\,.
\end{equation}
The first term describes the direct tunneling process from the $m$th quantized
torus to the chaotic sea with the rate $\gammad_m$, neglecting any influence
from nonlinear resonances. In the second term the coefficient
$A_{m,r}^{(r:s)}$ describes one resonance-assisted tunneling step from the $m$th
to the $(m$+$r)$th quantized torus via an $r$:$s$ resonance, while the factor
$\gammad_{m+r}$ describes the subsequent direct tunneling into the chaotic sea.
The last term accounts for two resonance-assisted and one direct tunneling step.
The three terms are visualized as insets in Fig.~\ref{fig:results_one_resonance}
in the $h$-regime where they are most relevant. 
Note, that the final step to the chaotic sea occurs due to direct 
tunneling, which is in contrast to previous studies 
\cite{EltSch2005,SheFisGuaReb2006}, where another resonance-assisted step is 
used.
We now discuss (i) the direct
tunneling rates $\gammad_{m}$, (ii) the coefficients $A_{m,r}^{(r:s)}$,
and (iii) the derivation of Eq.~\eqref{eq:rat_tunneling_rate}.

\begin{figure}[tb]
  \begin{center}
    \includegraphics[angle = 0, width = 85mm]{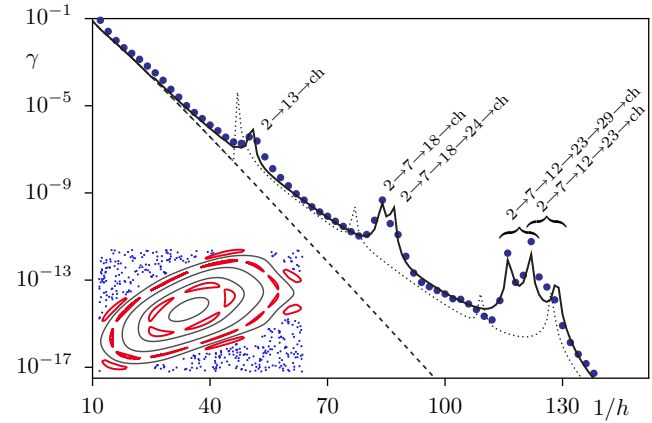}
    \caption{(Color online) Dynamical tunneling rates $\gamma$ from the quantized
             torus $m=2$ of a regular island with three dominant resonances.
             Numerical data (dots) is compared to the prediction of 
             Eq.~\eqref{eq:rat_tunneling_rate}
             (solid line). For comparison the direct tunneling contribution
             (dashed line, Eq.~\eqref{eq:gamma0}, Ref.~\cite{BaeKetLoeSch2008})
             and the result of Eq.~\eqref{eq:rat_tunneling_rate} without the 
             improvements (iia) and (iib) is presented (dotted line).
             The inset shows the regular island of the system \cite{RATMAPADD} 
             and the dominant $5$:$1$, $11$:$2$, and $6$:$1$ resonances which 
             cause tunneling steps as indicated by labels.
         }
    \label{fig:results_two_resonance}
  \end{center}
\end{figure}

\begin{figure}[tb]
  \begin{center}
    \includegraphics[angle = 0, width = 85mm]{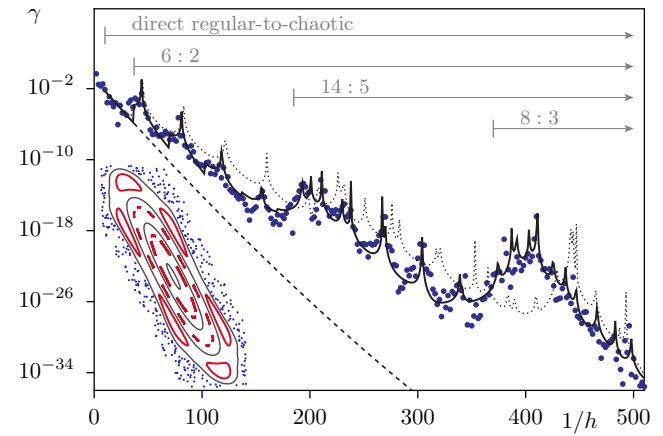}
    \caption{(Color online) Dynamical tunneling rates $\gamma$ from the
             innermost quantized torus $m=0$ of the regular island
             for the standard map at $K=3.5$.
             Numerical data (dots) is compared to the prediction of 
             Eq.~\eqref{eq:rat_tunneling_rate}
             (solid line). For comparison the direct tunneling contribution
             (dashed line, Eq.~\eqref{eq:gamma0}, Ref.~\cite{BaeKetLoeSch2008})
             and the result of Eq.~\eqref{eq:rat_tunneling_rate} without the 
             improvements (iia) and (iib) is presented (dotted line).
             The inset shows the island with the dominant
             $6$:$2$, $14$:$5$, and $8$:$3$ resonances and the arrows 
             indicate the regimes where they start to become relevant.
         }
    \label{fig:results_multi_resonance}
  \end{center}
\end{figure}

(i) The direct tunneling rates can be predicted
by an approach \cite{BaeKetLoeSch2008}
using a fictitious integrable system which
has to be chosen such that its classical dynamics
resembles the dynamics of the mixed system within the regular island as closely as possible.
The eigenstates $\psiregm$ of the fictitious integrable Hamiltonian $H_0$ are localized in the regular
region of $H$ and decay outside.
Using Fermi's golden rule, 
$\gammad_{m}\equiv\sum_{\text{ch}}|v_{\psiregm,\text{ch}}|^2$, in which
$v_{\psiregm,\text{ch}}=\langle\psich|U\psiregm$ couples 
$\psiregm$ to different chaotic states,
the direct tunneling rates are given as \cite{BaeKetLoeSch2008}
\begin{equation}
\label{eq:gamma0}
 \gammad_{m} = \Vert P_{\text{ch}}(U - \Ureg)
       \psiregm\Vert ^{2}
\end{equation}
where $\Ureg\equiv\exp(-\ui H_0 /\hbar)$ and $P_{\text{ch}}$ is a projector 
onto the chaotic region. Here we assume that there are no additional 
phase-space structures within the chaotic sea that affect the direct 
tunneling rates.
Note, that this approach is applicable for general systems with a mixed phase 
space \cite{BaeKetLoeRobVidHoeKuhSto2008,BaeKetLoeWieHen2009}. 
Different methods for the determination of the fictitious integrable 
system can be employed, based on the Lie transformation 
\cite{LicLie1983} (used in Figs.~\ref{fig:results_one_resonance} and 
\ref{fig:results_two_resonance}) or on the frequency map analysis 
\cite{LasFroCel1992} (used in Fig.~\ref{fig:results_multi_resonance}).
Details will be presented elsewhere \cite{BaeKetLoe:tbp}. 

(ii) The coefficients $A_{m,kr}^{(r:s)}$ in Eq.~\eqref{eq:rat_tunneling_rate}
depend on nonlinear resonances.
The classical dynamics of such an $r$:$s$ resonance is described in the 
corotating frame by the effective pendulum Hamiltonian 
\cite{BroSchUllcomb,LicLie1983}
\begin{equation}
\label{eq:pendulum_hamiltonian}
 \Hres(I,\theta) = H_0(I) -\Ors(I-\Irs) + 2\mathcal{V}_r(I)\cos(r\theta)
\end{equation}
with $\Ors=2\pi s/r$, where $2\mathcal{V}_r(I)\cos(r\theta)$ is the perturbation
in terms of the local action-angle variables $(I,\theta)$.
The simplest approach is the direct quantization 
of Eq.~\eqref{eq:pendulum_hamiltonian} in action-angle space by 
neglecting the action dependence of the coupling using 
$\Vrs\equiv \mathcal{V}_r(\Irs)$ and by making a quadratic approximation of 
$H_0(I)$ around the action $\Irs$ of the $r$:$s$ resonance, leading to 
$H_0(I)-\Ors(I-\Irs)\approx (I-\Irs)^2/2\mrs$. It is then found that the $r$:$s$
resonance couples the $m$th excited state $\psiregm$ 
of the resonance-free island to the state $\psiregmr$ with the
coefficient
\begin{equation}
\label{eq:Akr}
 A_{m,kr}^{(r:s)} = \frac{\Vrs \ue^{\ui\varphi}}{E_{m}-E_{m+kr}+kr \hbar\Ors}
\end{equation}
where $E_m = [I_m-\Irs]^2/(2\mrs)+\Ors I_m$ 
with $I_m \equiv \hbar(m+1/2)$
are the eigenvalues of the approximation of $H_0$.
$\Irs$, $\Vrs$, and $\mrs$ can be extracted from the classical
phase space \cite{EltSch2005,TomGriUll1995}.

(iii) To derive Eq.~\eqref{eq:rat_tunneling_rate} we start from 
Eq.~\eqref{eq:pendulum_hamiltonian} and apply quantum perturbation theory to 
obtain the perturbed regular states 
\begin{equation}
\label{eq:rat_perturbation_theory}
 |m\rangle = \psiregm + A_{m,r}^{(r:s)}\psiregmr +
                  A_{m,r}^{(r:s)}A_{m,2r}^{(r:s)}\psiregmtr + \dots
\end{equation}
These are now inserted into Fermi's golden rule
$\gamma_{m}\equiv\sum_{\text{ch}}|v_{|m\rangle,\text{ch}}|^2$ with
$v_{|m\rangle,\text{ch}}=\langle\psich|U|m\rangle$. 
Using the definition of $\gammad_{m}$ and
neglecting "off-diagonal" contributions of mixed 
terms (which at most amount to corrections of the order of a factor two)
leads to Eq.~\eqref{eq:rat_tunneling_rate}.

In the following we discuss two important improvements to the 
coefficients $A_{m,kr}^{(r:s)}$ in (ii):
(iia) On one hand, the quadratic approximation of $H_0(I)$ around $I=\Irs$
fails to predict the correct unperturbed energies $E_m$ especially for 
resonances close to the border of the regular island such that the peaks 
in the tunneling rates are located at a wrong position 
(see Figs.~\ref{fig:results_two_resonance} and \ref{fig:results_multi_resonance},
dotted lines). Here, we use a semiclassical procedure to
determine a \textit{global} approximation of the unperturbed system $H_0(I)$
\cite{BaeKetLoeSch2008}. Using its eigenenergies $\bar{E}_m$ in
Eq.~\eqref{eq:Akr} leads to improved peak positions in the tunneling rates.

(iib) 
On the other hand, the action dependence of the 
perturbation $\mathcal{V}_r(I)$ needs to be properly accounted for. 
Here we use the fact that the Hamiltonian $H(p,q,t)$ is analytic in the
``harmonic oscillator'' variables $(P,Q) \equiv \sqrt{2I}(-\sin \theta, 
\cos \theta)$ which result from $(p,q)$ through an analytical canonical
transformation.
This yields $\mathcal{V}_r(I) \propto I^{\frac{r}{2}}$ in lowest order in $I$ 
and $\mathcal{V}_r(I) e^{\pm i r \theta} \propto (Q \pm i P)^r$.
Quantization in $Q$-space then amounts to replacing $(Q+\ui P)/\sqrt{2 \hbar}$ 
by the ladder operator $\hat{a}$ satisfying $\hat{a} \psiregm = \sqrt{m}
\psiregmm$.
We therefore obtain in leading order
\begin{equation}
\label{eq:new_couplings}
 \tVrs^{m,kr} \equiv \psiregmpl\Hres\psiregmpr = 
  \Vrs\left(\frac{\hbar}{\Irs}\right)^{\frac{r}{2}}\sqrt{\frac{(m'+r)!}{m'!}}
\end{equation}
with $m'=m+(k-1)r$.
These new coupling parameters $\tVrs^{m,kr}$ depend on $\hbar$ and on
the quantum number $m$ of the coupled regular state. 
They replace $\Vrs$ in Eq.~\eqref{eq:Akr} and improve the tunneling 
rates in between the peaks.

We now predict dynamical tunneling rates with Eq.~\eqref{eq:rat_tunneling_rate}
using (i) the direct tunneling rates $\gammad_{m}$ from Eq.~\eqref{eq:gamma0}
describing the decay of the regular states by ignoring the influence of
resonances, (iia) the improved eigenenergies $\bar{E}_m$ of $H_0(I)$ 
and (iib) the action dependent couplings $\tVrs^{m,kr}$ from 
Eq.~\eqref{eq:new_couplings}. Fig.~\ref{fig:results_one_resonance} shows a 
comparison between this theoretical prediction (solid line) and numerical data 
(dots), obtained using absorbing boundary conditions in the chaotic sea.
In contrast to the sole application of direct tunneling with 
Eq.~\eqref{eq:gamma0} (dashed line) or resonance-assisted tunneling according 
to Ref.~\cite{EltSch2005} (gray line) we find excellent agreement
from the quantum to the semiclassical regime.
The phase space of the system is depicted in the inset of
Fig.~\ref{fig:results_one_resonance}. It shows one dominant $5$:$1$ resonance
inside the regular island. All other resonances are small and do not
contribute to the tunneling rates for the shown values of $h$.
We find that the direct tunneling rates $\gammad$ following from
Eq.~\eqref{eq:gamma0} explain the numerical data for
$1/h \lesssim 20$. For larger $1/h$ the $5$:$1$ resonance is important and 
gives rise to two characteristic peaks corresponding to the
coupling of the ground state $m=0$ to $m=5$ and $m=10$. Our theoretical
prediction excellently reproduces the peak positions and heights.

As a next example we consider a situation where three dominant resonances exist,
namely a $5$:$1$ resonance near the center of the regular island, an
$11$:$2$, and an outer $6$:$1$ resonance, see the 
inset in Fig.~\ref{fig:results_two_resonance}.
Here, multiresonance couplings occur and the tunneling rates from the torus 
$m=2$ are determined with Eq.~\eqref{eq:rat_tunneling_rate} by a summation over
all relevant $r$:$s$ resonances. For example the peak at $1/h \approx 84$
in the tunneling rates is caused by the coupling of the state $m=2$ to
the $18$th excited state via the $5$:$1$ and the $11$:$2$ resonance.
In Fig.~\ref{fig:results_two_resonance} we compare numerical data (dots) to the
prediction of Eq.~\eqref{eq:rat_tunneling_rate} and find very good agreement.
For comparison we also show the rates (dotted line) that result from 
using Eq.~\eqref{eq:rat_tunneling_rate} together with Eq.~\eqref{eq:Akr}
without the improvements (iia) and (iib).
We still find overall agreement to the average decrease 
of the tunneling rates with $1/h$. However, the reproduction of the 
positions and heights of the individual peaks requires to use
a global approximation for the unperturbed energies (iia) and to take into 
account the action dependence of the coupling matrix elements (iib).

The paradigmatic model of quantum chaos is the standard map [$T(p)=p^2/2$,
$V(q)=-K/(4\pi^2)\cos(2\pi q)$]. For $K=3.5$ it has a large generic
regular island with a dominant $6$:$2$ resonance and further relevant
$14$:$5$ and $8$:$3$ resonances. These are the largest 
low-order resonances. Fig.~\ref{fig:results_multi_resonance}
shows the comparison between numerical data and the prediction of 
Eq.~\eqref{eq:rat_tunneling_rate} far into
the semiclassical limit. We find quantitative
agreement over $30$ orders of magnitude in $\gamma$.
The direct tunneling process is relevant for $1/h \lesssim 30$ beyond
which several regimes for the tunneling rates caused by different resonances
are identified. We attribute the
small deviations below the prediction of Eq.~\eqref{eq:rat_tunneling_rate}
to destructive interference of different tunneling sequences leading to the 
same final state.
 
An important question is when the resonances become relevant, i.e.\ where is
the transition between the direct tunneling regime $h \lesssim \A$, where 
resonances play no role, and the resonance-assisted tunneling regime $h\ll \A$.
The latter requires that the $(m$+$r)$th quantized torus exists inside the island,
i.e.\
\begin{equation}
 \frac{\A}{h} \geq m+r+\frac{1}{2}.
\end{equation}
The position $\hpeak$ of the first peak in $\gamma_m$ arises when Eq.~\eqref{eq:Akr}
diverges for
$E_{m}-E_{m+r}+r \hbar\Ors=0$. Using the quadratic approximation for $H_0(I)$,
we find
\begin{equation}
\label{eq:comp_peak}
 \frac{\Ars}{\hpeak} = m + \frac{r}{2} + \frac{1}{2}
\end{equation}
with $\Ars \equiv 2\pi \Irs$. The first peak appears when the $r$:$s$
resonance encloses $m$+$r/2$ quantized tori.
However, the influence of the resonance may appear much earlier than $1/\hpeak$.
We define the transition point $1/\hres$ when the
first two terms in Eq.~\eqref{eq:rat_tunneling_rate} are equal. 
Using $\sqrt{2\mrs\Vrs}=\dArs/16$ \cite{EltSch2005}, where $\dArs$ is the area 
enclosed between the separatrices of the $r$:$s$ resonance, we obtain 
\begin{equation}
\label{eq:comp_crit}
 \frac{\dArs}{r\hres}\,\frac{\dArs}{\Ars}\,\sqrt{\frac{\gammad_{m+r}(\hres)}{\gammad_m(\hres)}}\,
  \frac{1}{\left(\hres/\hpeak-1\right)} = \frac{128}{\pi^2}.
\end{equation}
This criterion explicitly involves the size $\dArs$ of the $r$:$s$ resonance 
chain measured with respect to $r$ Planck cells and to the area $\Ars$
enclosed by the resonance, as well as the ratio of the direct
tunneling rates from the $(m$+$r)$th and the $m$th quantized torus.
As the left-hand side of Eq.~\eqref{eq:comp_crit} is expected to display a 
monotonous increase with $1/\hres$ towards the singularity at $1/\hpeak$, 
resonance chains with a large area $\dArs$ will lead to a lower transition
point $1/\hres$ at which the crossover from direct to resonance-assisted
tunneling appears. While, e.g., the first peak of the $5$:$1$ resonance
in Fig.~\ref{fig:results_two_resonance} appears at $1/\hpeak \approx 130$ it
dominates the tunneling process already at $1/\hres \approx 40$ which is even before the
contributions from other resonances set in.
Equations~\eqref{eq:comp_peak} and \eqref{eq:comp_crit} confirm 
the intuition that low-order resonances with small $r$,$s$ and 
large $\Delta \Ars$ are most relevant.

In conclusion, we have combined an improved resonance-assisted tunneling theory
with the theory of direct tunneling using a fictitious integrable system
to predict dynamical tunneling rates from the quantum regime, $h\lesssim \A$, to the
semiclassical regime, $h \ll \A$.
Excellent quantitative agreement with numerically determined tunneling rates
is found on the level of individual resonance peaks, which emphasizes
the validity of the underlying direct and resonance-assisted mechanisms.
We therefore expect that these mechanisms leave their characteristic traces in
semiclassical approaches based on complex classical trajectories
\cite{ShuIkecomb}, and allow one to understand and predict
tunneling rates also in more complex, multidimensional quantum systems.

We thank S.~Keshavamurthy for useful discussions and 
the DFG for support within the Forschergruppe 760
``Scattering Systems with Complex Dynamics''.

\end{document}